\documentclass[aps,showpacs,showkeys,11pt]{revtex4}

\usepackage{amssymb, lineno}

\begin{document}
\title{Exposed-key weakness of $\alpha \eta$}
\author{Charlene Ahn}
\email{cahn@toyon.com}
\affiliation{
Toyon Research Corporation,
6800 Cortona Drive,
Goleta, CA 93117}
\author{Kevin Birnbaum}
\email{Kevin.M.Birnbaum@jpl.nasa.gov}
\affiliation{
JPL M/S 161-135,
4800 Oak Grove Drive, Pasadena, CA 91109
}

\begin{abstract}
The $\alpha \eta$ protocol given by Barbosa \emph{et al.}, PRL 90, 227901
(2003) claims to be a secure way of encrypting messages using mesoscopic
coherent states.  We show that transmission under $\alpha \eta$ exposes
information about the secret key to an eavesdropper, and we estimate the
rate at which an eavesdropper can learn about the key.  We also consider
the consequences of using further randomization to protect the key and how
our analysis applies to this case.  We conclude that $\alpha \eta$ is not
informationally secure.
\end{abstract}

\maketitle

Encryption of sensitive data is an ubiquitous problem in military,
commercial, and even personal communications. Quantum mechanics can be
used to solve this problem by generating a key that can be proven to be
unconditionally secure via the BB84 protocol (e.g., \cite{shor-preskill});
this key can then be securely used in a one-time pad. However, BB84 is
difficult to implement and has relatively low bit rates compared to
current data transmission rates \cite{hughes:fiber-QKD}. To combat these
disadvantages, another quantum encryption scheme which can send encrypted
data at very high rates and which is easily implemented has been proposed
in \cite{barbosa:mesoscopic-security}. This protocol is often called the
$\alpha \eta$ protocol or the $Y-00$ protocol and purports to draw its
security from confusing an eavesdropper Eve by using the uncertainty on
any measurement she can make.

The $\alpha \eta$ protocol has not been shown to be unconditionally secure
as has BB84; indeed, the attack given in \cite{lo:cryptographic-attacks} (see also \cite{Yuen-Lo_response})
shows that at best the security of $\alpha \eta$ must be complexity-based,
as are current classical ciphers. In this paper, we show that $\alpha
\eta$ has an additional disadvantage that current classical ciphers do not
have: transmission of the ``encrypted'' states actually leaks information
about the key to an eavesdropper, even if that eavesdropper has no
information about the message. Such a weakness has been independently
described briefly in \cite{Yuen-random}, and in more depth in
\cite{Donnet-attack} and \cite{Yuen-Donnet-response}; here, we calculate
information loss and estimate a bound on the efficacy of explicit attacks
such as \cite{Donnet-attack}. In the remainder of this paper, we will
describe the $\alpha \eta$ scheme, show that in practice there is no
advantage created for Bob over Eve via quantum limits on measurement, and
estimate how much information Eve can learn about the key from Alice's
transmission. We will also discuss a technique given in \cite{Yuen-KCQ}
and \cite{Yuen-random}, Deliberate Signal Randomization, and how our
analysis applies there.

 In the $\alpha\eta$ protocol, the symbols transmitted from Alice to Bob are physically encoded as mesoscopic coherent states (mean photon numbers $N \sim 10-10^5$) of varying phase or polarization.  Without loss of generality, we will consider the states to have the same polarization and amplitude, and varying phase.  We will take the number of symbols to be $M$, and the code states to be $|\alpha(j)\rangle =| e^{2\pi i j/M}\sqrt{N}\rangle$, where $j\in\{0,\dots M-1\}$.  The states to be sent are selected by the following technique: 
 \begin{enumerate}
 \item Starting with a key $K$ of size $L$ bits, use a pseudo-random number generator to produce a running key $R$.  In order to send a message of size $Q$ bits, $Q\mathrm{log}(M/2)$ bits of $R$ must be computed. 
 \item Take $\mathrm{log}(M/2)$ bits of $R$, $k_q\in\{0,\dots M/2-1\}$.  Take one bit of message, $b_q\in\{0,1\}$.
 \item Compute $j_q = ((k_q \mathrm{mod} 2)\oplus b_q)M/2 + k_q$.  Alice sends the corresponding state $|\alpha(j_q)\rangle$.  This may be interpreted as a ``basis'' of $k_q$, which corresponds to an angle within a half-circle, and a ``signal'' of $(k_q \mathrm{mod} 2)\oplus b_q$ which determines whether the upper or lower half-circle is used.
 \item Repeat steps $2$ and $3$ using successive bits of the message and strings of the running key, until the entire message has been sent (a total of $Q$ times).
 \end{enumerate}
 Note that knowledge of $\{j_q\}$ unambiguously determines the message bits $\{b_q\}$, regardless of the key.  The protocol relies upon the condition that the code states are not perfectly distinguishable for any physical receiver.  This can be guaranteed if $M/\sqrt{N}$ is sufficiently large because the quantum states of neighboring symbols will have high overlap.
 
 There is much discussion in the literature of ``advantage creation.''  The principle of this is that if the intended receiver, Bob, knows the secret key, then he can do an optimal measurement to distinguish the two possible states Alice may send (depending on the message bit).  An eavesdropper, Eve, who does not know the key, must discriminate more possible states, and hence must perform a different measurement.  This measurement will be non-optimal at distinguishing the two states which may actually be sent, and therefore it is claimed that Eve will necessarily have a higher bit-error rate than Bob.  However, this comparison is fair only if Bob and Eve are both given a signal with equal amplitude.  In practice, if Bob is receiving a signal which has been attenuated by more than 3dB, or has been put through an amplifier (amplifiers reduce signal to noise by at least 3dB), then it is possible that Eve has received more signal than Bob.  Furthermore, the optimal measurement that Bob could perform (the Dolinar receiver \cite{JM-detection}) has not been experimentally demonstrated, although work on this is pending \cite{JM-private}, so instead it is proposed that Bob perform a homodyne measurement (where the local oscillator phase is determined by the running key).  For a given signal strength, however, homodyne has only a weak advantage over the demonstrated adaptive-phase technique (see \cite{armen}), and only a 3dB advantage over the commonplace heterodyne measurement.  That is, if the attenuation factor is more than $2/3$ (more than 5dB loss), then it is possible that Eve has twice the signal that Bob has, and even if she only performs heterodyne, she actually has a measurement advantage over Bob (assuming he does homodyne).  We therefore believe that for practical implementations, Bob will not have a measurement advantage over Eve, and the quantum aspects of the problem can be modeled by a classical system with appropriate noise. 
 
We will treat the system classically by assuming that Alice
 computes $j_q$, then sends
\begin{equation}
\label{eqn:class}
j_q' = (j_q + w_q)\mathrm{mod} M.
\end{equation}
Here $w_q$ is a gaussian-distributed random variable, with mean zero
and standard deviation $\sigma\geq M/(4\pi\sqrt{N}).$ The equality is held for an ideal phase measurement with unit efficiency.  Imperfect detection and loss will introduce an attenuation $\eta$ of the signal, such that 
\begin{equation}
\label{eqn:sigma}
\sigma = M/(4\pi\sqrt{\eta N}).
\end{equation}
For example, if in the path from Alice to Bob the beam goes through a long fiber with 10dB of attenuation, and then is detected by heterodyne with 80\% quantum efficiency, then $\eta_{Bob} = \eta_{loss}\eta_{het}\eta_{q.e.} = (0.1)(0.5)(0.8)=0.04$.  The approximation of gaussian distributed phase noise is good for $\eta N \gg 1$.  We will assume that the $w_q$ ``sent'' to Bob and Eve are uncorrelated.

Here is a brief derivation of the formula for $\sigma$ given in Eq.~\ref{eqn:sigma}.  The light pulse which encodes each symbol is in a mode with annihilation operator $\hat{a}$.  We define quadrature components $\hat{Q}_{\phi} = e^{i\phi}\hat{a}^\dag + e^{-i\phi}\hat{a}$, $\hat{P}_{\phi} = i(e^{i\phi}\hat{a}^\dag - e^{-i\phi}\hat{a})$.  These operators have the commutation relation $[\hat{Q}_{\phi},\hat{P}_{\phi}]=2i$.  A coherent state has minimum uncertainty on the quadratures, and thus has $\langle\Delta\hat{Q}_{\phi}^2\rangle = \langle\Delta\hat{P}_{\phi}^2\rangle = 1$.  For coherent states with amplitude large compared to unity, a phase measurement can be approximated as a measurement of a quadrature which is $\pi/2$ with respect to the coherent state amplitude.  That is, for a coherent state $|e^{i\theta}\sqrt{N}\rangle$, we pick quadratures $\hat{Q}_{\theta},\hat{P}_{\theta}$, and measure $\hat{P}_{\theta}$.  Our estimate of the phase is given by $\theta' = \theta + P'_{\theta}/\langle\hat{Q}_{\theta}\rangle$, where $P'_{\theta}$ is the result of our measurement.  Our estimator $\theta'$ is unbiased and minimum variance in the limit of $N\gg 1$.  Notice that we have used a $\mathrm{tan}(x) = x$ approximation, and that we would have to know both the phase and amplitude of the state in advance in order to actually do this ideal measurement on  a single state.  The variance of $\theta'$ is given by $\langle \hat{P}_{\theta}^2\rangle/\langle\hat{Q}_{\theta}\rangle^2 = 1/(4N)$.  Therefore the standard deviation of $j' = \theta'M/2\pi$ is $M/(4\pi\sqrt{N})$.

It is claimed that the $\alpha\eta$ scheme is secure when $\sigma \gg 1$, because then Eve cannot accurately infer $j_q$ from $j_q'$.  The claim is that Eve will estimate the message bit as $$b_q'^{(Eve)} = (j_q'\mathrm{mod}2)\oplus(j_q'\mathrm{mod}\frac{M}{2}).$$  This estimate will have a very high error rate because $j_q'\mathrm{mod}2$ has a very low correlation with $j_q\mathrm{mod}2$ due to the noise term.  Meanwhile, it is claimed Bob will have a low error rate because he will use his knowledge of $k_q$ to compute
\begin{equation}
	x_q' = \left\{
		\begin{array}{ll}
			0 & \textrm{if $|k_q - j'_q| < M/4$ or $|k_q - j'_q| > 3 M / 4$}\\
			1 & \textrm{otherwise}
		\end{array}
		\right\}
\end{equation}
That is, $x_q'$ is zero when $j_q'$ is in the same
half-plane as $k_q$, and one when it is in the other half-plane.  Bob will
then estimate the message as $b_q'^{(Bob)} = x_q' \oplus
(k_q\mathrm{mod}2)$, which will have a low error rate if $\sigma\ll M$.

However, the claims of security do not consider that the record of all of the measurements $\{j_q'\}$ \emph{do} give some information on $\{j_q\}$, from which Eve can obtain information on $\{k_q\}$, and ultimately on $K$.  Once Eve knows $K$, she can then compute all $\{k_q\}$, and hence can decrypt in the same way as Bob.

We will now estimate the information gain on $j_q$ from a measurement $j_q'$, in the limit $M\gg\sigma\gg 1$ (the limit where it is claimed there is good security against Eve and a low error rate for Bob).  The initial entropy on $j_q$ before the measurement is $H_0$, which we will take as
\begin{equation}
H_0 = -\sum_{m=0}^{M-1}p_m \mathrm{log}(p_m) = \mathrm{log}(M).
\end{equation}
We have assumed no prior knowledge of the symbol, i.e.~uniform probabilities $p_m=1/M$ for all $M$.  Without loss of generality, we will take the actual symbol prepared by Alice to be $j_q = M/2$.  Then Eve's probabilities for the symbol are
\begin{equation}
p_m \simeq \frac{1}{\sigma\sqrt{2\pi}} \int_{m-1/2}^{m+1/2}\mathrm{exp}\left(-\frac{(x-\frac{M}{2})^2}{2\sigma^2}\right)dx \approx \frac{1}{\sigma\sqrt{2\pi}}\mathrm{exp}\left(-\frac{(m-\frac{M}{2})^2}{2\sigma^2}\right).
\end{equation}
Then the entropy after the measurement is 
\begin{eqnarray}
H_1 &\approx& -\int_0^M \frac{1}{\sigma\sqrt{2\pi}}\mathrm{exp}\left(-\frac{(m-\frac{M}{2})^2}{2\sigma^2}\right)\left( -\mathrm{log}(\sigma\sqrt{2\pi})-\frac{(m-\frac{M}{2})^2}{2\sigma^2}\mathrm{log}(e) \right)dm\nonumber\\
&\approx& \mathrm{log}(\sigma\sqrt{2\pi e}).
\end{eqnarray}
Therefore, the information gained is 
\begin{equation}
\label{eqn:infogain}
H_0 - H_1 \approx \mathrm{log}(\frac{M}{\sigma\sqrt{2\pi e}}) =
\mathrm{log}(2\sqrt{\frac{2\pi}{e}\eta N})\approx
\frac{1}{2}\mathrm{log}(\eta N)+ 1.6
\end{equation} 
bits for each symbol that Eve measures.  This is the information gained on $j_q$; the information gained on $k_q$ is approximately $1$ bit less, because the message bit obscures $1$ bit of the running key per symbol.  Since $K$ and all $\{ k_q \}$ are deterministically related, in principle information on $k_q$ can be converted to information on $K$.  We therefore take $U = \mathrm{log}(\sqrt{\frac{2\pi}{e}\eta N})$ as an upper bound on Eve's information on $K$ per measured symbol.  

We expect Eve's information to grow linearly with the number of symbols
because the use of a pseudo-random number generator implies that values of
$K$ which have similar values of $k_q$ will have uncorrelated values of
$k_{s}$ for $s\neq q$.  In other words, the pseudo-random number generator
will redistribute Eve's prior probabilities back to the flat distribution
for each new symbol.  This approximation will of course break down when
the Eve's entropy on $K$ is low, such that her entropy on the key will
only asymptotically approach zero as the number of symbols goes to
infinity.  In this latter limit, the prior probabilities will be strongly
peaked, and additional measurements of $k_q$ will provide little
additional information on $K$.  Eve's entropy on the key will transition
from linear decline to asymptotic decay after measuring approximately
$n_0=L/U$ symbols, by analogy to the unicity distance \cite{shannon1949}
of a classical deterministic cipher used to encode a redundant (reduced
entropy) message. We note that this unicity distance is very similar to
the unicity bound calculated in \cite{Yuen-random}.

We estimate from our unicity bound that Eve may have enough information to determine the key with high probability when $Q \gg n_0$.  Let us take an example by considering the experimental demonstration of $\alpha \eta$ given in \cite{Yuen-fiber}. In that demonstration, Alice and Bob share a key $K$ with $L=4400$ bits, and Alice sends states with $N=40000$ photons. Let us now assume that Eve detects with total efficiency $\eta^{(Eve)}=0.1$.  Then for each symbol Alice sends, Eve gains about $7.6$ bits of information, or about $U \approx 6.6$ bits of information about the key.  Since each symbol transmits $1$ bit of information to Bob, then we can see that if Alice sends much more than $n_0 = 4400/6.6\sim 668$ bits to Bob, then Eve will have enough information to find the key, and hence decrypt all of messages that were sent with that key.

It is important to note that in the above, we did not assume that Eve initially had any information on the message.  This is not a plaintext attack; this weakness simply comes from the fact that the symbols $\{j_q\}$ contain information on the key which is not totally obscured by the noise.  

For comparison, let us consider a simple additive streaming cipher.  We will define this cipher by the following procedure:
 \begin{enumerate}
 \item Starting with a key $K$, use a pseudo-random number generator to produce a running key $R$.  In order to send a message of size $Q$ bits, $Q$ bits of $R$ must be computed. 
 \item Take one bit of $R$, $k_q\in\{0,1\}$.  Take one bit of message, $b_q\in\{0,1\}$.
 \item Alice sends $j_q = k_q \oplus b_q$ over a noiseless channel. 
 \item Repeat steps $2$ and $3$ on successive bits of the message and running key, until the entire message has been sent (a total of $Q$ times).
 \end{enumerate}
 If Eve's entropy on the message is initially $H^{Eve}(message) = Q$ (no plaintext is known), then  her entropy of the key does not decrease.  Essentially, the key encrypts the data, and the data encrypts the key.  On the other hand, if Eve starts with some knowledge of the message, she can perform a known plaintext attack on the key which will succeed with high probability if $Q-H^{Eve}(message))\gg L$ \cite{shannon1949}.   That is, every bit of the message which is known to Eve can be used to reveal one bit of the key.  The $\alpha\eta$ protocol has a similar property, in that if one bit of the message $b_q$ is known, then knowledge of $j_q'$ can more effectively be used to find $k_q$.  We estimate that in the presence of known plaintext, Eve can determine $K$ with high probability when $Q (U+1)-H^{Eve}(message)) \gg L$.  Thus, from an information-theory standpoint, the $\alpha\eta$ protocol is worse than the simple additive stream cipher.  
 
Ref. \cite{Donnet-attack} gives an explicit attack exploiting this leak in
information security for linear feedback shift register (LFSR) based
stream ciphers. Figure 6 of their paper graphs the minimal number of symbols
needed for a successful attack, $S_0$, as a function of $g = L$, the number of
bits about the generator that need to be learned. Our analysis simply
gives $$S_0 \gg \frac{g}{U}.$$ That is, the relationship between $g$ and
$S_0$ is linear, which is roughly what \cite{Donnet-attack} finds in
Figure 6.  Using Eq.~\ref{eqn:infogain}, with $\alpha = 2\sqrt{\eta N} =
300$, and subtracting a bit for information about the message gives the bound $S_0
\gg g/7.8$, to be compared with the numerical result of the attack
employed in \cite{Donnet-attack} of $S_0\approx 40g$.  Of course, we are
assuming optimal use of information, which is not necessarily achieved in a
practical attack.

It may be claimed that while $\alpha\eta$ does not have security in the
information theory sense, it may have complexity based security, in that
it would take unreasonably large computational resources for Eve to
convert her information on $\{j_q'\}$ into information on $K$.  An
analysis of the computational complexity of this task would depend on the
choice of pseudo-random number generators, and is beyond the scope of this
paper. 

Independently, a similar observation regarding the exposure of the key has
been recently made by H.~Yuen in \cite{Yuen-Donnet-response} (and see also
\cite{Yuen-random} and \cite{Donnet-attack}). In
\cite{Yuen-random,Yuen-Donnet-response}, it is argued that a technique that
is called Deliberate Signal Randomization (DSR) will serve to add
information-theoretic security to the key. This technique simply involves
Alice sending a random state in the half-plane around the state she would
send under the non-DSR version given above. That is to say, equation
(\ref{eqn:class}) would become
\begin{equation}
\label{eqn:dsr}
j_q' = (j_q + w_q + \beta_q)\mathrm{mod} M,
\end{equation}
where $\beta_q$ is a uniformly distributed random variable between
$-M/4$ and $M/4$.

Under such a system, it is true that Eve will not learn anything about
the key, but at the expense of introducing error into the transmission
when $\beta_q$ is close to $-M/4$ or $M/4$.  (If the distribution on
$\beta_q$ is truncated or otherwise changed to lessen the error, it
should be clear how to modify the above calculation to show that
information is still being given to Eve.)  
Ref. \cite{Yuen-Nishioka04_response} calculates that this error would
be about one percent. Now, the message could conceivably be encoded in
an error-correcting code, or some low-entropy message, such as one in
English, could be sent with this system, thus allowing correction of
the errors by Bob. However, this introduces redundancy into the message, which
decreases Eve's entropy on the message, which she can exploit to collect information about the
key as above. We conclude that the use of DSR could conceivably lessen the
breach of information-theoretic security, but that the $\alpha \eta$
scheme with DSR is still informationally insecure, and gives more information to an attacker than a simple classical additive stream cipher.

It is clear that in order to avoid errors, or the redundancy necessitated by an error-correcting code, one should set the variance of $w_q$ in Eq.~\ref{eqn:dsr} to zero.  In this case, where any measurement noise is negligible, it is possible to have good information-theoretic security.  However, in terms of information this is equivalent to the simple additive streaming cipher described earlier.  Although there are $M$ possible symbols in the channel, precisely one bit is conveyed in each transmission due to the random $\beta$ term.  The transmitted bit is computed from the message and the key.  The exact nature of the computation is different, but the information contained is the same.  

An additional problem with DSR, also noted briefly in
\cite{Yuen-Donnet-response}, is that it introduces another source of
randomness with a particular distribution that must be fed into the system
at a very high data rate. This rather negates the spirit of the original
proposal, which depended on the fast generation of randomness given by the
measurement of a coherent state. Practically, as well, finding true random
sources at high data rate is difficult, and using a pseudo-random number
generator leads to information-theoretic exposure of the second generator
by the same logic as that used above.

Another advantage of $\alpha \eta$ stated in \cite{Yuen-random} is
that it gains some security due to the physical nature of the states
being sent: it may in practice be difficult to perform the measurements needed to eavesdrop on a channel with this encoding.  However, since an effective eavesdropping strategy is to employ a heterodyne or dual-homodyne measurement, we do not see a great difference in the practical difficulty of eavesdropping from the difficulty of the legitimate receiver, or from the receiver of any coherent communication system.

In conclusion, we have found that the $\alpha \eta$ protocol does not have good
information-theoretic security.  While the information-theoretic security is not always the primary concern, we believe it is an important factor in the assessment of a cryptosystem.

This work has been supported by the Air Force Office of Scientific
Research (AFOSR) under Phase I Small Business Technology Transfer program grant
FA9550-05-C-0091.

A portion of this research was carried out at the Jet Propulsion Laboratory, California Institute of Technology, and was sponsored by AFOSR through an agreement with the National Aeronautics and Space Administration.
\bibliographystyle{elsart-num}
\bibliography{qkd_sttr}

\end{document}